\documentclass[sigconf, 10pt, nonacm]{acmart}  

\AtBeginDocument{%
  \providecommand\BibTeX{{%
    Bib\TeX}}}

\usepackage{balance}
\usepackage{graphicx}
\usepackage{textcomp}
\usepackage{xcolor}
\usepackage{float}
\usepackage{algorithm}
\usepackage{algpseudocode}
\usepackage{caption}
\usepackage{subcaption}
\usepackage{makecell}
\usepackage{enumitem}
\def\BibTeX{{\rm B\kern-.05em{\sc i\kern-.025em b}\kern-.08em
    T\kern-.1667em\lower.7ex\hbox{E}\kern-.125emX}}

\begin{document}


\title{How Much Sensing is Enough? Landslide Detection from Sparse and Noisy 3D Observations}

\settopmatter{authorsperrow=3}    

\author{Arijit Pal}
\affiliation{
\small\href{mailto:arijitpalop@gmail.com}{arijitpalop@gmail.com}
\small\href{https://orcid.org/0009-0008-0214-9155}{0009-0008-0214-9155}\\
\institution{IIEST Shibpur\textsuperscript{\#}}\city{}\country{}}

\author{Abhijit Mohanty}
\affiliation{
\small\href{mailto:abhijitmohanty782@gmail.com}{abhijitmohanty782@gmail.com}
\small\href{https://orcid.org/0009-0009-6238-4592}{0009-0009-6238-4592}\\
\institution{IIEST Shibpur\textsuperscript{\#}}\city{}\country{}}

\author{Dipanjyoti Paul}
\affiliation{
\small\href{mailto:dipanjyoti@ai.iitkgp.ac.in}{dipanjyoti@ai.iitkgp.ac.in}
\small\href{https://orcid.org/0000-0001-9079-7524}{0000-0001-9079-7524}\\
\institution{IIT Kharagpur\textsuperscript{\#\#}}\city{}\country{}}

\author{Sandip Chakraborty}
\affiliation{
\small\href{mailto:sandipc@cse.iitkgp.ac.in}{sandipc@cse.iitkgp.ac.in}
\small\href{https://orcid.org/0000-0003-3531-968X}{0000-0003-3531-968X}\\
\institution{IIT Kharagpur\textsuperscript{\#\#}}\city{}\country{}}

\author{Basabdatta Palit}
\affiliation{
\small\href{mailto: basabdatta@it.iiests.ac.in}{ basabdatta@it.iiests.ac.in}
\small\href{https://orcid.org/0000-0003-0834-2307}{0000-0003-0834-2307}\\
\institution{IIEST Shibpur\textsuperscript{\#}}\city{}\country{}}

\renewcommand{\shortauthors}{Pal et al.}


\begin{abstract}
Dense 3D terrain information is critical for accurate landslide detection; however, acquiring such data at large scale is often limited by sensing infrastructure, coverage, and environmental conditions. Emerging sensing modalities (e.g., mmWave, low-cost radar, sparse LiDAR) can operate under adverse conditions, but inherently produce sparse and noisy 3D observations, raising an important question: \emph{how much sensing is sufficient for reliable landslide detection?} Given the absence of large-scale sparse-terrain datasets, we have constructed a physics-informed dataset by combining digital elevation models from OpenTopography with soil moisture data from NASA's SMAP. Subsequently, we have randomly generated landslide events and introduced controlled sparsification and stochastic noise to emulate sensing constraints. We have then developed a learning-based framework to reconstruct the underlying terrain and identify landslide-prone regions using a CNN + XGBoost method. Our results show that with only 20\% of the original observations, the proposed method achieves a detection accuracy of 80\% and a ROC-AUC of 0.868, with a reconstruction error (L2 Chamfer Distance) of 99.67~$m^2$ on the test set. We have further studied the impact of sparsity, highlighting the trade-off between sensing density and detection performance.
\end{abstract}



\begin{teaserfigure}
  \footnotesize
  \textsuperscript{\#}Indian Institute of Engineering Science and Technology (IIEST), Shibpur, Howrah, West Bengal 711103, India.\\
  \textsuperscript{\#\#}Indian Institute of Technology Kharagpur, Kharagpur, West Bengal 721302, India.
\end{teaserfigure}

\maketitle

\section{Introduction}
Landslides and subsidence events pose significant threats to human life, infrastructure, and environmental stability, particularly in regions of steep terrain, high rainfall, and rapid urbanization. Their accurate and timely detection requires continuous monitoring of terrain elevation, slope, soil moisture, and surface deformation. Conventional approaches include airborne remote sensing methods such as multispectral imaging and Synthetic Aperture Radar (SAR)~\cite{LandSubsidenceGNSSInSar,Nature_LandSlide_Casagli2023}, and ground-based techniques such as LiDAR and interferometry~\cite{Kumar2020Jharia}. However, satellite-based methods have fixed acquisition frequency and limited adaptability, while optical and LiDAR-based systems are sensitive to fog, rainfall, and low visibility. Furthermore, existing approaches~\cite{wang2026landslide} typically assume either dense 3D observations or coarse-grained 2D representations. High-resolution monitoring, therefore, requires dense sensing infrastructure, which is often prohibitively expensive for continuous, large-area deployment.

Emerging sensing modalities, such as mmWave radar~\cite{Hao2025} and sparse LiDAR~\cite{SparseLiDAR2024}, on the other hand, offer improved robustness under adverse environmental conditions at lower deployment costs. Although they have been used for Point Cloud Data (PCD) reconstruction~\cite{milipcd2022}, their 3D resolution is often sparse and noisy due to coverage gaps and measurement uncertainty. Consequently, landslide detection with these modalities must be performed on partial and noisy terrain representations. To the best of our knowledge, existing works neither address landslide detection under such conditions nor quantify the impact of sensing sparsity on detection performance. This motivates us to address the question: \textit{how much sensing is sufficient for reliable landslide detection?}

In this work, we have developed a framework for detecting landslides from sparse and noisy data. In the absence of large-scale datasets of sparse terrain observations, we have first constructed a physics-informed synthetic dataset of landslides in hilly terrain using Digital Elevation Models (DEMs) obtained from OpenTopography~\cite{opentopography}. Although DEMs enable computation of features such as slope, curvature, and flow accumulation~\cite{nasa_dem,guth2021dem}, they do not represent a true 3D point cloud, as each spatial location corresponds to a single elevation value. So, the representations generated in this work are pseudo-3D point clouds only. We have incorporated soil moisture from the NASA SMAP dataset~\cite{smap_nasa}. We have then generated landslide events stochastically based on a slope and moisture-dependent instability criteria, followed by controlled terrain deformation to obtain the ground truth labels. To emulate sensing and deployment constraints, we have next introduced controlled sparsification and Gaussian noise to yield incomplete and noisy 3D terrain observations.

Subsequently, we have designed a learning-based pipeline to reconstruct the underlying 3D terrain from these observations. A coarse-grained surface is first constructed from the noisy sensor data. It is then refined through multi-modal fusion of the noisy and coarse-grained PCDs, guided by the Chamfer distance between the ground truth and the reconstructed images. The reconstructed images serve as inputs to a classification model that combines a CNN,  with XGBoost over the combined feature space. CNN is used to captures local spatial patterns from elevation patches, while XGBoost models non-linear relationships . Therefore, unlike existing works that assume high-quality inputs, our framework explicitly considers degraded sensing conditions and evaluates landslide detection as a function of sensing sparsity. The results demonstrate that the proposed method achieves $80$\% detection accuracy using only $20$\% of the original observations.
The contributions of this work are, thus, as follows:
\begin{enumerate}
  \item a physics-guided DEM-based synthetic dataset of landslides with (a) a slope-moisture-driven landslide generation, (b) sparsified and perturbed by noise injection to emulate the sparse data obtained from resource-constrained sensors,
  \item an ML-based 3D PCD reconstruction pipeline leveraging Chamfer loss between the ground truth and the reconstructed PCDs, using multi-modal fusion of noisy and coarse-grained PCDs,
  \item a landslide-detection mechanism based on a CNN-XGBoost model trained on the reconstructed 3D PCDs.
\end{enumerate}
\section{Synthetic Landslide Dataset Generation}
\subsection{Landslide-prone Terrain Generation from DEM}\label{sec:terr_gen}
We have first constructed a continuous terrain representation from DEM data and derived the geomorphological and hydrological features required for landslide simulation. The details follow.
\label{sec:surfacecreation}
\begin{enumerate}[leftmargin=*, itemsep=0pt, topsep=0pt]
    \item The geographic coordinates $P = \{\mathtt{(longitude, latitude, z)}\}$ of the input DEM are transformed into local Cartesian coordinates $(x,y)$ using a first-order equirectangular approximation, centered at the region's mean latitude~\cite{Snyder1987}.
    \item The scattered points are next converted to a continuous terrain surface using Delaunay triangulation with linear interpolation~\cite{Virtanen2020}, discretized into a structured grid of size $G_{size} \times G_{size}$. Missing values near the grid boundaries are filled using nearest-neighbour interpolation.
    \item The elevation grid thus obtained is first smoothed using a Gaussian filter with standard deviation $\sigma = 2$ grid cells to suppress high-frequency noise~\cite{lindsay2019}. Terrain features, such as slope, aspect, and profile and plan curvature, are then extracted using finite difference methods. The Terrain Ruggedness Index (TRI) is calculated as the sum of the absolute differences between the elevation of the central cell and those of its neighbouring cells, considering an $8$-connected neighborhood. Roughness is calculated as the difference between the maximum and minimum elevation in the same neighborhood.
    \item To inject soil moisture, we have used the regional average soil moisture from the NASA-SMAP dataset~\cite{smap_nasa} as the base $M_{base}$. In the absence of fine-grained soil moisture measurements, we have assumed a linear function to model soil moisture as:
\begin{equation}
M = M_{base} + \alpha \cdot TWI_{norm},
\end{equation}
where $\alpha$ controls the influence of topographic convergence and $TWI_{norm}$ is the normalized Topographic Wetness Index (TWI). We set $\alpha=0.8$ as a representative value to capture moderate-to-high sensitivity of moisture to topographic accumulation. TWI is computed from water flow accumulation on the elevation grid using a deterministic eight-direction (D8) flow routing scheme~\cite{riihimaki2021,winzeler2022}.
    \item The grid-based features are mapped back to the original point set $P$ through interpolation, so that each point is associated with its terrain attributes for the subsequent stages of landslide simulation and sensing degradation.
\end{enumerate}

\subsection{Landslide Generation and Ground Truth Construction}\label{sec:landslide_gen}
\subsubsection{\textbf{Risk Region Identification}}
In this work, potential landslide-prone regions are identified as those with slope greater than $28^\circ$, soil moisture exceeding $0.25$, and elevation within the top $25\%$. These thresholds capture relatively steep, saturated, and elevated locations. Furthermore, we have retained only the top $0.4\%$ of the points as landslide epicenters, so that landslides remain localized events.

\subsubsection{\textbf{Landslide Simulation via Terrain Deformation}}
Landslide events are simulated by introducing randomized localized depressions at the selected epicenters. The deformation is modeled using a Gaussian function:
\begin{equation}
\Delta z(x,y) = d \cdot \exp \left( -\frac{(x - h_x)^2 + (y - h_y)^2}{2\sigma_L^2} \right),
\end{equation}
where $(h_x, h_y)$ denotes the epicenter coordinates, $d$ controls the depth of deformation, and $\sigma$ determines the spatial extent. To introduce variability in landslide size and intensity, $d$ is drawn uniformly from $1 \pm 0.5$ m and the radius from $300 \pm 50$ m for each epicenter, with $\sigma_L$ set to half of the radius. The depressions of overlapping epicenters are superposed.

The deformation field $\Delta z(x,y)$ is defined as a positive downward displacement with the modified elevation surface:
\begin{equation}
z'(x,y) = z(x,y) - \Delta z(x,y).
\end{equation}
This formulation thus captures erosion-dominated landslides. The modified elevation field is mapped back to the original point set to obtain the updated terrain values at each location.

\subsubsection{\textbf{Ground Truth Label Generation}}
Ground truth labels for landslides are generated by comparing the elevation before and after deformation, by thresholding $|\Delta z|$ as: $
L =1, \text{if } |\Delta z| > \tau,
L=0, \text{otherwise}$.
 $\tau = 1\,\text{m}$ is used to distinguish significant terrain failure from minor variations.

\subsection{Sparse and Noisy Observation Generation}\label{sec:noiseyPCD}
To emulate low-resolution sensing, the point coordinates of the terrain representation obtained in Section~\ref{sec:landslide_gen} are first discretized to a fixed spatial resolution ($0.1$ m), and the duplicate points within the same spatial cell are removed. To reflect the limited field-of-view of practical sensing systems, the terrain is then partitioned into independent spatial tiles of fixed size ($1200 \times 1200$ m), and tiles with insufficient data are discarded. Within each tile, the point cloud is thinned by randomly retaining $15\%$, $20\%$, or $25\%$ of the points.

To account for measurement uncertainty, each remaining point $(x,y,z)$ is then displaced by independent zero-mean Gaussian noise with a standard deviation of $0.5$ m per axis, clipped at $\pm1.5$ m. The noise is intended to represent generic errors in range and angle estimation.
\section{Terrain Reconstruction from Sparse and Noisy images and Landslide Detection}

\subsection{The Dense Point Cloud Reconstruction}
\begin{figure}[t]
    \centering
    \includegraphics[width=1\linewidth]{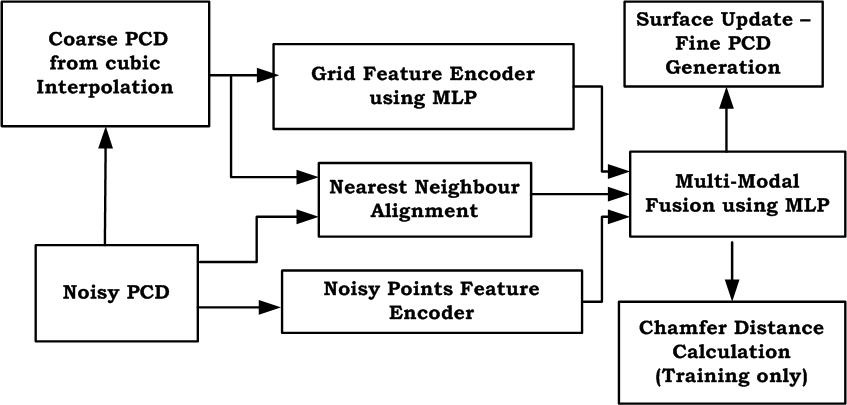}
    \vspace{-6mm}
    \caption{Model Architecture for generating a Fine PCD from the noisy PCD.}
    \label{fig:ReconstructionModel}
\end{figure}
The sparse and noisy pseudo point clouds generated in Section~\ref{sec:noiseyPCD} consist of irregularly spaced 3D points $P_n = (x_i, y_i, z_i)$. These do not provide sufficient spatial coverage to compute terrain features for landslide detection. We, therefore, reconstruct a dense pseudo point cloud that approximates the underlying terrain surface, as outlined in Fig.~\ref{fig:ReconstructionModel}. This step is indispensable when operating with real measured data.

We first define a structured grid $G = \{(x_j, y_j)\}$ over the 2D spatial domain. For grid locations coinciding with observed points, the elevation values are directly assigned. For the remaining locations, elevations are estimated using cubic interpolation, resulting in an initial continuous surface $Z_G = \{z_j\}$. As this surface does not fully capture local variations in sparse regions, a nearest-neighbor alignment is established between each grid location $(x_j, y_j)$ and its closest observed point in $P_n$. The surface is then refined using a multi-modal fusion of features implemented via a multi-layer perceptron (MLP). Features derived from the grid-based representation and the associated observed points of the noisy PCD are combined to predict point-wise corrections $\Delta z_j$, which are applied to obtain the reconstructed point cloud. The model is trained to minimize the Chamfer distance between the reconstructed point set $P_r$ and the ground truth $P_{gt}$~\cite{milipcd2022}:
\begin{equation}
\mathcal{L}_{CD}(P_r, P_{gt}) =
\frac{1}{|P_r|} \sum_{x \in P_r} \min_{y \in P_{gt}} \|x - y\|^2 +
\frac{1}{|P_{gt}|} \sum_{y \in P_{gt}} \min_{x \in P_r} \|x - y\|^2.
\label{eq:chamfer_d}
\end{equation}
\begin{figure}[t]
    \centering
    \includegraphics[width=\linewidth]{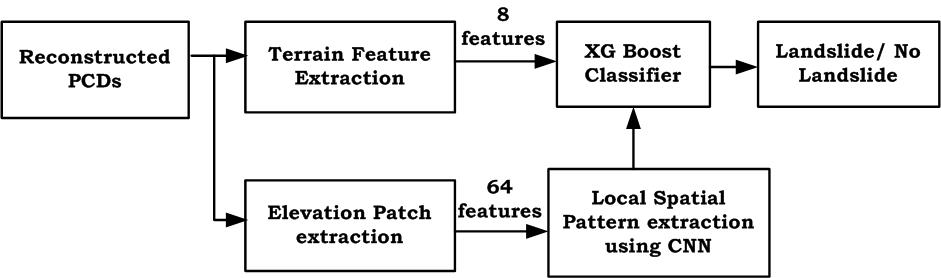}
    \vspace{-6mm}
    \caption{The pipeline for Landslide Detection from Reconstructed PCD. }
    \label{fig:PredictionModel}
\end{figure}
\begin{figure*}[htbp]
    \centering
    \begin{subfigure}[b]{0.32\textwidth}
        \centering
        \includegraphics[
            width=0.8\linewidth,
            trim=0 1 0 0, 
            clip
        ]{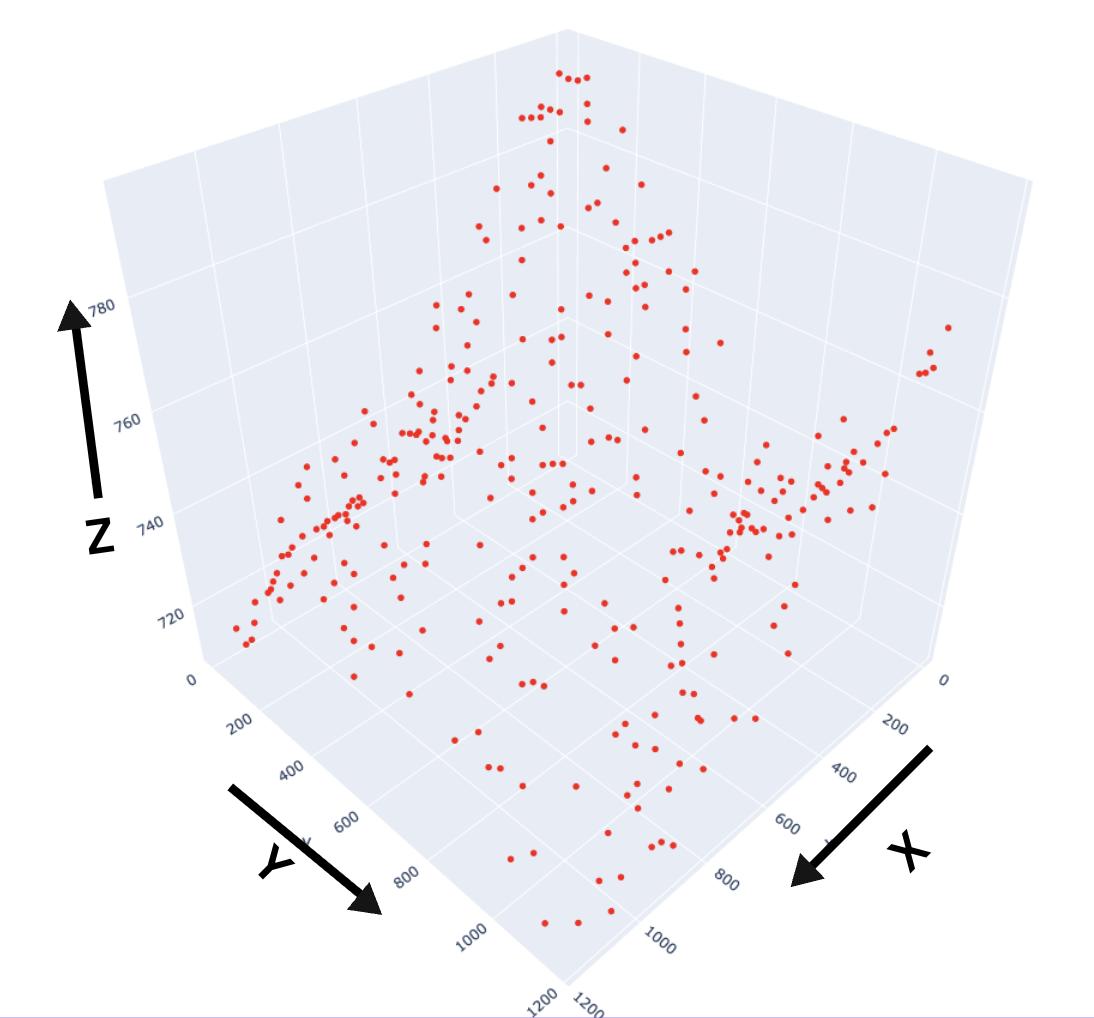}
        \caption{Noisy Input}
        \label{fig:noisy}
    \end{subfigure}
    \hspace*{0.2cm}
    \begin{subfigure}[b]{0.32\textwidth}
        \centering
        \includegraphics[
            width=0.8\linewidth,
            trim=0 1 0 0, 
            clip
        ]{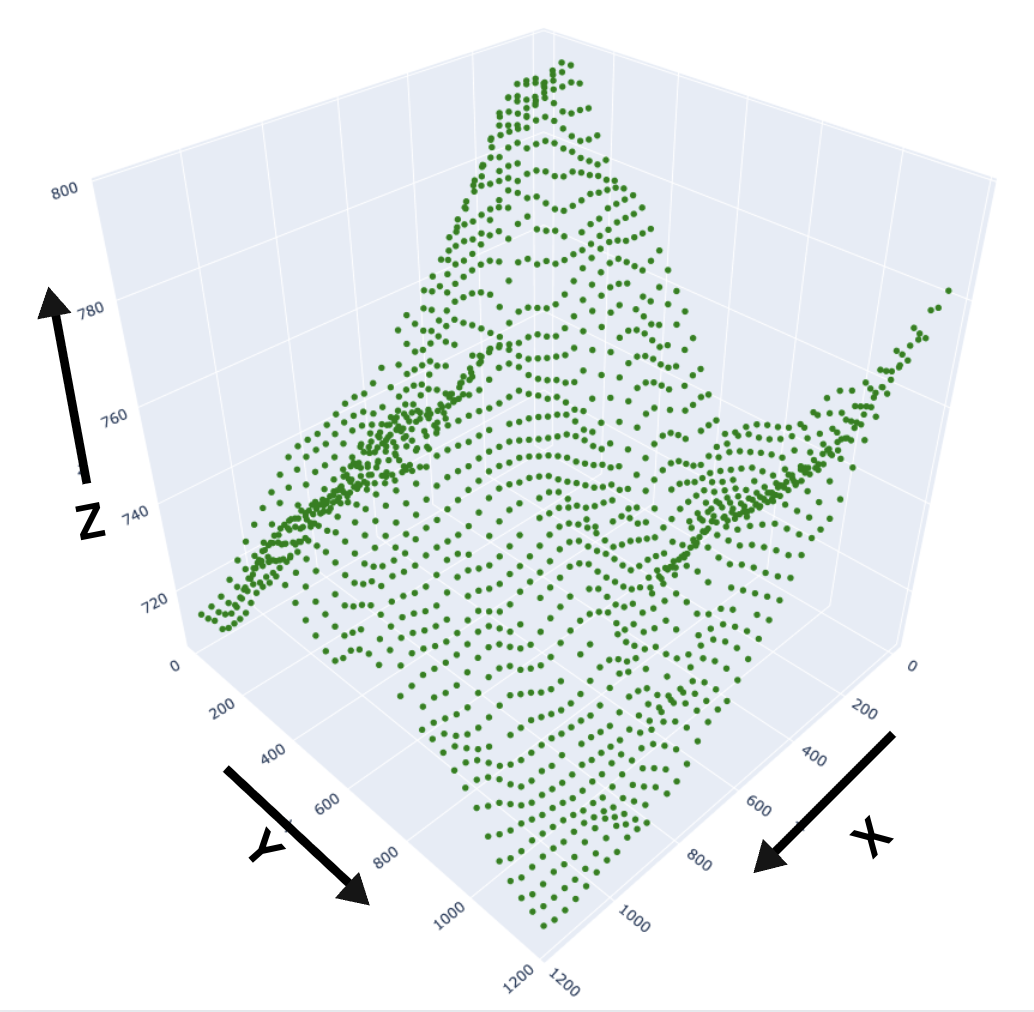}
        \caption{Ground Truth}
        \label{fig:gt}
    \end{subfigure}
    \hspace*{0.2cm}
    \begin{subfigure}[b]{0.32\textwidth}
        \centering
        \includegraphics[
            width=0.8\linewidth,
            trim=0 0 0 0, 
            clip
        ]{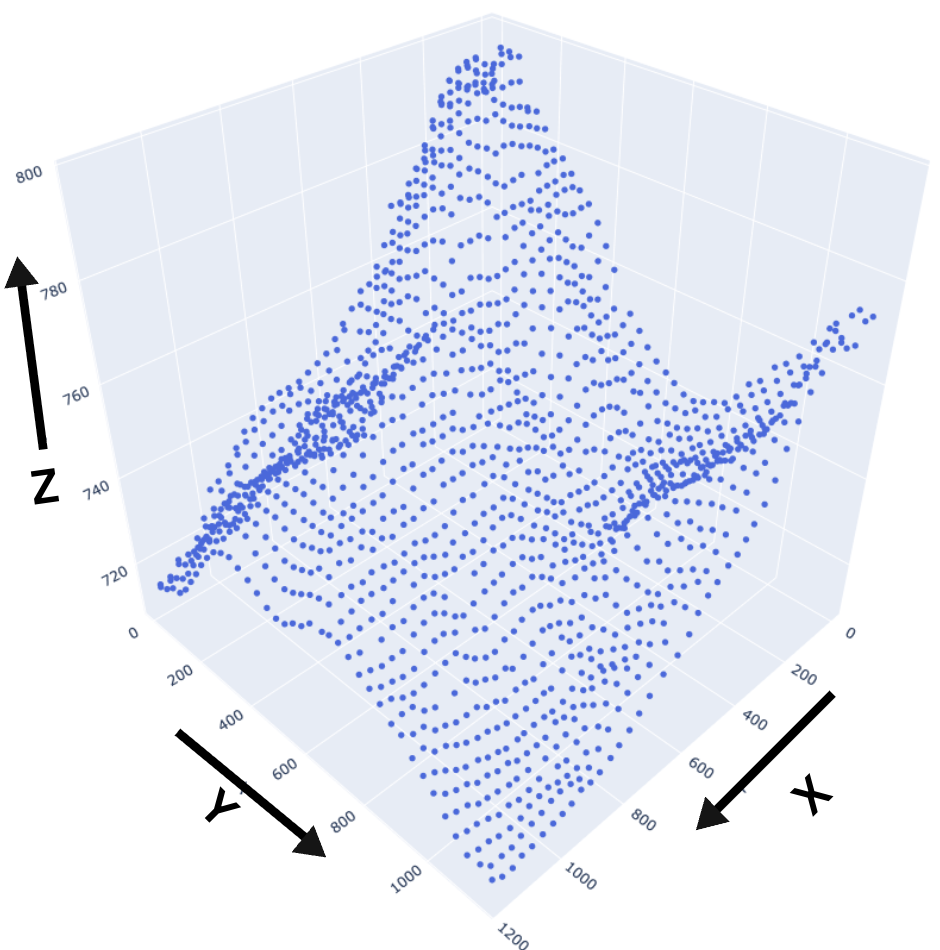}
        \caption{Reconstructed Output}
        \label{fig:refined}
    \end{subfigure}
    \caption{Comparison of Point Clouds:
    (a) Noisy input, (b) Ground truth, (c) Fine PCD obtained after reconstruction.}
    \label{fig:results_comparison}
\end{figure*}
\vspace*{-0.3cm}
\subsection{Landslide Detection Framework}\label{sec:landslide_det}
After reconstruction, landslide detection is carried out on the reconstructed terrain, as outlined in Fig.~\ref{fig:PredictionModel}. For each reconstructed point, the label is assigned by nearest-neighbor matching with the ground truth point cloud.

Terrain features such as slope, aspect, profile and plan curvature, TRI, and TWI are computed from the reconstructed surface. Slope and aspect are obtained from first-order derivatives, while profile and plan curvature are obtained from second-order derivatives of the surface (Section~\ref{sec:surfacecreation}); the profile and plan curvature capture the variation of slope along and perpendicular to the direction of maximum gradient, respectively. The TRI is computed over a $3 \times 3$ neighborhood as in Section~\ref{sec:surfacecreation}. The TWI is calculated as $\text{TWI} = \ln\left(\frac{A}{\tan(\beta)}\right)$~\cite{winzeler2022} for an area $A$ and a local slope angle $\beta$. Soil moisture is computed as in Section~\ref{sec:terr_gen}.

These features describe the overall shape of the terrain, but not its fine-scale surface geometry.
To capture the latter, we have used a convolutional neural network (CNN) on local elevation patches.
For each point, a fixed-size patch centred on that point is cropped from the reconstructed surface and normalized.
The CNN maps this patch to a compact embedding of the local surface structure.
Subsequently, the embedding is concatenated with the terrain features and given to an XGBoost classifier.
Therefore, the classifier can model the non-linear relationships between the physically derived features and the learned local structure.
\vspace*{-0.3cm}
\section{Results and Discussion}
\subsection{PCD Reconstruction Model}
\textbf{Experimental Setup:} The reconstruction model was evaluated on $1350$ pairs of noisy terrain tiles ($1.2 \times 1.2$ km, picked randomly) and ground truth, obtained from the NASA-SRTM DEMs of three regions, (30\textdegree N, 78\textdegree E), (15\textdegree N, 74\textdegree E), and (10\textdegree N, 77\textdegree E), each covering 591~${km}^{2}$ with 675,000 points.  The tile pairs were randomly split into 70\% for training, 20\% for validation, and 10\% for testing. As adjacent tiles overlap slightly at the edges, the split is not strictly geographically disjoint.  The model was trained with a batch size of 4 for 100 epochs. The results of this section consider 20\% retained points.

\textbf{Quantitative Results:} We have evaluated the reconstruction quality using the symmetric L2 Chamfer Distance (CD) as in eq.~\eqref{eq:chamfer_d}. On the test set, the model achieves a CD of 99.67~$m^2$, i.e., $\sqrt{\text{CD}} \approx 10$ m, which is $<1\%$ of the tile edge (1200 m). The validation loss reduces from $121.35~m^2$ (0.92\%) at the initial stage to $109.66~m^2$ (0.87\%) at convergence, while the training loss decreases from $141.50~m^2$ (0.99\%) to $106.44~m^2$ (0.86\%), with the learning rate decreasing from $5.0 \times 10^{-4}$ to $3.1 \times 10^{-5}$ under cosine annealing. The closeness of the two losses indicates that the model generalizes well to unseen terrain tiles.  Fig.~\ref{fig:results_comparison} shows that the refinement effectively smooths out and preserves the global structure of the terrain.

\vspace*{-0.3cm}
\subsection{Landslide Detection Model}
\textbf{Dataset and Training Setup:} The landslide detection model is trained on approximately 2 million points across the three regions mentioned earlier. Of these around 1.51 million are non-landslide regions and 0.51 million are landslide regions. The points are randomly split into 80\% for training (of which 15\% is held out for validation) and 20\% for testing. The inputs are the terrain features of Section~\ref{sec:landslide_det} and a $32 \times 32$ normalized (zero mean and unit variance) elevation patch around each point. The patches are passed through a CNN having - (1) four convolutional layers (16, 32, 64, 64 filters; $3\times3$ kernels) with batch normalization and ReLU activation, (2) a max-pooling layer after the second block, (3) adaptive average pooling, and (4) a fully connected layer producing a 64-dimensional embedding. The CNN is trained using the Adam optimizer with a learning rate of $10^{-3}$, cosine annealing, and a weighted cross-entropy loss, with weights proportional to the ratio of non-landslide to landslide samples. The XGBoost classifier is trained with 500 estimators, a maximum depth of 6, and a learning rate of 0.05.

\begin{figure}[htbp]
    \centering
    \begin{subfigure}[b]{0.48\linewidth}
        \centering
        \includegraphics[
            width=1.15\linewidth,
            trim=2 0 0 0,
            clip
        ]{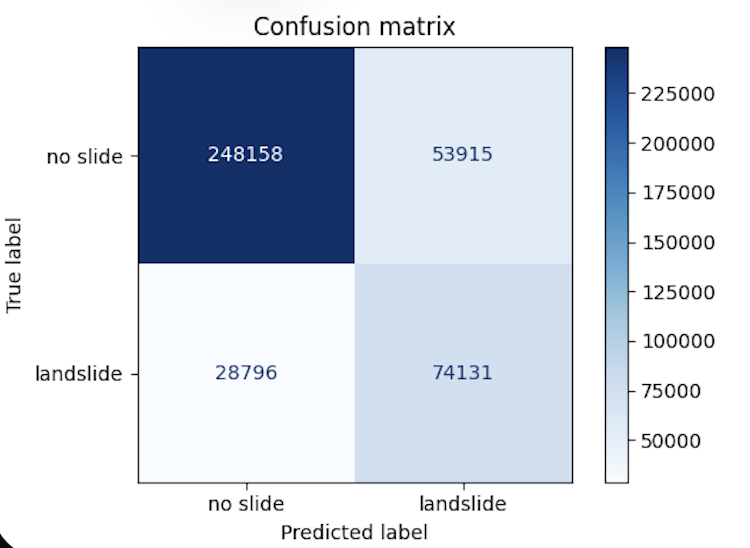}
        \caption{Confusion Matrix}
        \label{fig:confusionmatrix}
    \end{subfigure}
    \hfill
    \begin{subfigure}[b]{0.48\linewidth}
        \centering
        \includegraphics[
            width=0.85\linewidth,
            trim=0 0 0 0,
            clip
        ]{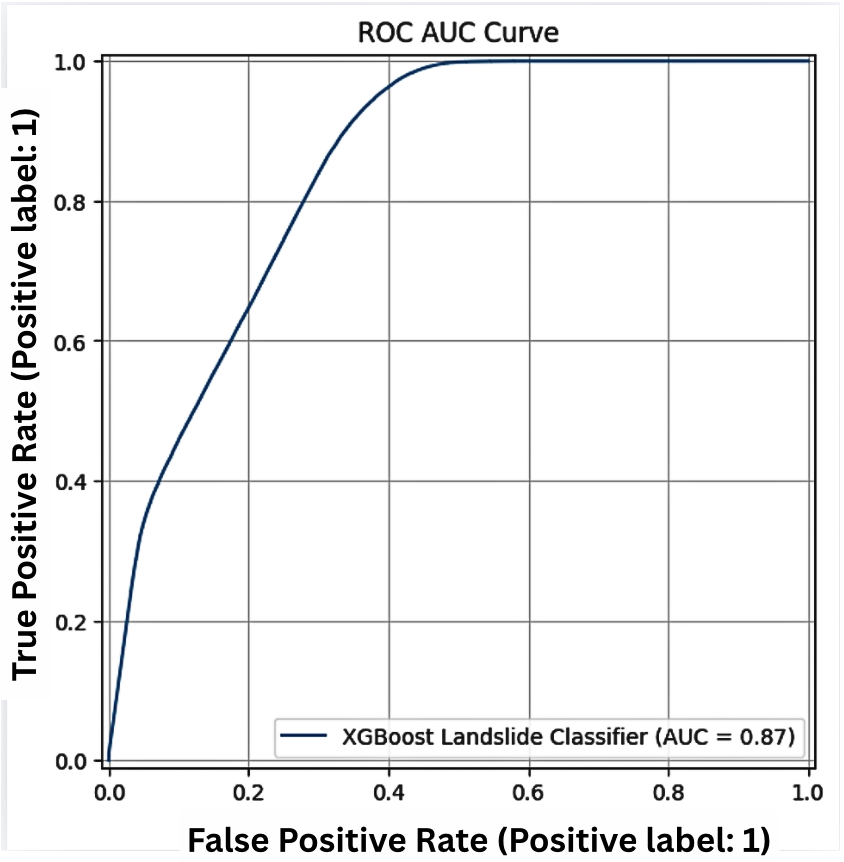}
        \caption{ROC-AUC Curve}
        \label{fig:rocauc}
    \end{subfigure}
\vspace{-3mm}
    \caption{(a) Confusion matrix, (b) ROC-AUC curve of landslide detection for 20\% available points..}
    \label{fig:performance_plots}
\end{figure}

\textbf{Classification Results:} The proposed framework achieves an overall accuracy of 80\% and a ROC-AUC of 0.868 (Fig.~\ref{fig:rocauc}). For the landslide class, the model achieves a recall of 0.72 and a precision of 0.58, resulting in an F1-score of 0.64, while for the non-landslide class, the F1-score is 0.86. Landslide points constitute only about 25\% of the data, and class-weighted training favors recall over precision. Hence, most landslide events are detected, at the cost of a higher false-alarm rate (Fig.~\ref{fig:confusionmatrix}), which is a reasonable trade-off for early-warning applications. As the split is performed at the point level, neighboring points and their patches may fall in different splits, and the accuracy should therefore be read as an in-distribution estimate; evaluation on geographically held-out regions is left for future work.

\subsection{Effect of the Retention Ratio}
Table~\ref{table:comparison} summarizes the effect of the fraction of retained points on both reconstruction and detection performance. As the fraction increases, the Chamfer distance decreases, as expected. The detection accuracy, however, saturates at $20\%$ ($80\%$, against $79.5\%$ at $25\%$), and a difference of $0.5$ percentage points is within the variation that may be expected across training runs, which we have not measured. Hence, within the range studied ($15$--$25\%$), a limited number of points appears sufficient for reliable detection, while the behavior outside this range remains unexplored.
\begin{table}[htbp]
\caption{Effect of the Retention Ratio}
\vspace{-4mm}
\begin{center}
\begin{tabular}{|l|c|c|}
\hline
\makecell{\textbf{Retained} \\ \textbf{Points}} 
& \makecell{\textbf{L2 Chamfer} \\ \textbf{distance (${m}^{2}$)}} 
& \makecell{\textbf{Landslide Prediction} \\ \textbf{Accuracy}}\\
\hline
15\% & 116.31 (0.90\%) & 75.6\% \\
\hline
20\% & 109.66 (0.87\%) & 80\% \\
\hline
25\% & 104.23 (0.85\%) & 79.5\% \\
\hline
\end{tabular}
\label{table:comparison}
\end{center}
\vspace{-4mm}
\end{table}
\section{Conclusion and Future Work}
In this work, we looked at landslide detection under sparse and noisy sensing conditions. The proposed pipeline combines terrain reconstruction with a CNN and XGBoost-based landslide detection model. By using both terrain features and local spatial patterns, the model is able to achieve an accuracy of 80\% and a ROC-AUC of 0.868 even when only partial observations are available. The results show that, for the retention ratios studied (15–25\%), a limited number of points suffices for reliable detection on synthetic data.

A limitation of the current study is that it is based on synthetic data: landslides are simulated as Gaussian depressions placed using slope, moisture, and elevation rules, which are also classifier features, so part of the reported accuracy may reflect the labeling rules; the noise is generic Gaussian noise rather than a calibrated sensor model; and the train/test splits are random rather than region-wise. So, in future our framework will be tested on real-world sparse sensor data, with region-wise evaluation and ablations of each module. Furthermore, in the future, we also plan to include and analyze temporal information to capture gradual terrain changes before a landslide.
{\small
\setlength{\bibsep}{0pt}
\bibliographystyle{acm}
\bibliography{LandSense_ICDCN_Camera_Ready}
}
\end{document}